# Gravity induced effective photon mass in the presence of gravity gradient

**Igor I. Smolyaninov**
Saltenna LLC, 1751 Pinnacle Drive, Suite 600 McLean VA 22102-4903 USA

**Email**:
 igor.smolyaninov@saltenna.com

**Abstract**

Geometry and gravity induced effective photon mass is known to arise in many cases, such as various optical waveguides, Kaluza-Klein theories, and many other optical and general relativity situations. Here we study the appearance of effective photon mass in the Newtonian limit due to the presence of gravity gradient in a locally inertial reference frame. The effective photon mass squared appears to be proportional to the local gravity gradient, and it becomes tachyonic around unstable equilibrium locations. A similar effect is observed in the Kottler–Møller spacetime where the absolute value of the gravity-induced tachyonic effective photon mass appears to coincide with the Unruh temperature. We demonstrate that similar to Unruh effect, a bath of thermal radiation is observed near the unstable equilibrium, which temperature is defined by the local gravity gradient, and which remains unchanged in the c→∞ limit.

## Introduction

Both simple geometrical constraints and various non-trivial spacetime geometries are known to induce effective photon mass in many situations. For example, it is well known that photons propagating inside various optical waveguides aquire effective mass [1,2]. A geometry-induced effective photon mass also arises in Kaluza-Klein theories [3,4] and numerous other general relativity situations [5], even though the electromagnetic field equations remain massless at the fundamental level. For example, the dispersion law of photons propagating inside an empty rectangular optical waveguide looks like dispersion law of a massive quasi-particle:

$$\frac{\omega^2}{c^2} = k_z^2 + \frac{\pi^2 I^2}{d^2} + \frac{\pi^2 J^2}{b^2} \quad , \tag{1}$$

where $\omega$ is the photon frequency, $k_z$ is the photon wave vector in the z-direction, $d$ and $b$ are the waveguide dimensions, and $I$ and $J$ are the mode numbers in the x- and y- directions, respectively [2]. The effective inertial rest mass of the photon in the waveguide (which is equal to its effective gravitational mass) is

$$m_{eff} = \frac{\hbar \omega_{ij}}{c^2} = \frac{\hbar}{c}\left(\frac{\pi^2 I^2}{d^2} + \frac{\pi^2 J^2}{b^2}\right)^{1/2} \tag{2}$$

Note that the paricle masses in Kaluza-Klein theories have exactly the same origin [3,4]. The particle masses are defined by the scale of microscopic hidden ("waveguide") dimensions.

Moreover, recent advances in transformation optics highlighted the now well-understood fact that many electromagnetic and general relativity situations are interchangeable [6,7], so that empty but curved spacetime manifolds may appear as effective dielectric media. Indeed, in an empty but possibly curved spacetime, the effective dielectric permittivity and magnetic permeability are [6,7]:

$$\varepsilon^{ij} = \mu^{ij} = -\frac{\sqrt{-g}}{g_{00}} g^{ij} \tag{3}$$

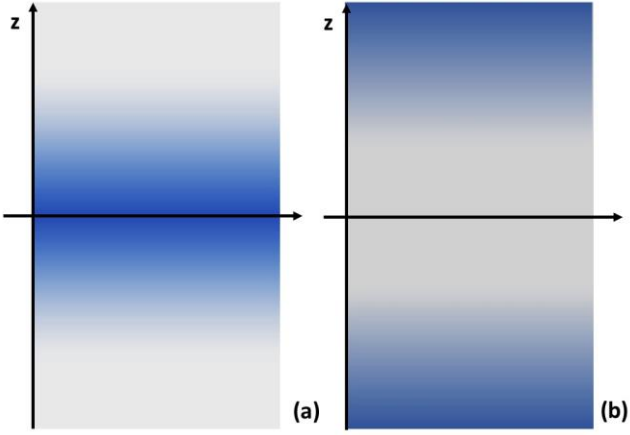

**Fig. 1**. Schematic representation of the effective dielectric permittivity $\varepsilon$ and magnetic permeability $\mu$ in the stable (a) and unstable (b) electromagnetic waveguides, which are equivalent to the Newtonian spacetime metric (4) in the case of $H>0$ (a) and $H<0$ (b), respectively. The relative magnitudes of $\varepsilon$ and $\mu$ are represented by halftones.

As a result, the electromagnetic waveguide geometries which are known to induce effective photon mass may find direct analogues in general relativity.

In this paper we consider an effective electromagnetic medium which is equivalent to the Newtonian limit of a locally inertial reference frame, which is perturbed by a gravity gradient. It appears that such a spacetime geometry looks similar to a dielectric waveguide, in which photons aquire effective mass. We will demonstrate that the so determined effective photon mass squared is proportional to the local gravity gradient, and it becomes tachyonic around unstable equilibrium points. Moreover, a similar effect is observed in the Kottler–Møller spacetime where the absolute value of the gravity-induced effective tachyonic photon mass appears to coincide with the Unruh temperature. We will discuss particle production by gravity gradients in the unstable equilibrium case and relate it to the appearance of horizons in phase space.

## Effective photon mass in the Newtonian limit

Let us consider the Newtonian limit of general relativity in which the spacetime metric is defined as [7]

$$ds^2 = -\left(1+\frac{2\phi}{c^2}\right)c^2 dt^2 + \left(1-\frac{2\phi}{c^2}\right)\left(dx^2 + dy^2 + dz^2\right) \quad , \quad (4)$$

where $\phi$ is the weak gravitational potential. For the sake of simplicity, we will assume that the gravitational field at $z=0$ is zero, and a gravity gradient exists only in $z$ direction:

$$\phi = Hz^2 \quad , \quad (5)$$

so that the reference frame is inertial in the vicinity of $z=0$ (the easiest way to realize such a spacetime metric in the experiment is to use a reference frame which is co-moving with a mass attached to a spring oscillating in the $z$ direction). Based on Eq.(3), the effective electromagnetic medium representing metric (4) should have the following $z$-dependent permittivity and permeability:

$$\varepsilon = \mu = \frac{\left(1-\frac{2\phi}{c^2}\right)^{3/2}}{\left(1+\frac{2\phi}{c^2}\right)^{1/2}} \approx 1 - \frac{4\phi}{c^2} = 1 - \frac{4Hz^2}{c^2}$$
(6)

It is easy to observe that at $H>0$ such a medium behaves as a planar gradient electromagnetic waveguide, as illustrated in Fig. 1a. The electromagnetic waves inside such a waveguide are guided by an elevated refractive index $n=(\varepsilon\mu)^{1/2}$ in the middle plane of the waveguide. As described in detail in [8], the wave equation inside such a gradient waveguide may be written as

$$-\frac{\partial^2\psi}{\partial z^2} + \left(-\frac{\varepsilon(z)\mu(z)\omega^2}{c^2} - \frac{1}{2}\frac{\partial^2\varepsilon}{\varepsilon\partial z^2} + \frac{3}{4}\frac{(\partial\varepsilon/\partial z)^2}{\varepsilon^2}\right)\psi = -k^2\psi \quad , \quad (7)$$

where $k$ is the photon wave vector in the $xy$ plane, and the effective wave function is introduced as $\psi=\varepsilon^{1/2}E_z$. After simple transformations this wave equation may be re-written in the vicinity of $z=0$ as

$$-\frac{\partial^2\psi}{\partial z^2} + \left(-\frac{\omega^2}{c^2} + \frac{4H}{c^2}\right)\psi = -k^2\psi \quad (8)$$

In this wave equation the $4H/c^2$ term plays the role of effective mass squared. The effective photon mass is positive at $H>0$, which is understandable based on similarity between this situation and a regular optical waveguide [2].

On the other hand, at $H<0$ the wave equation becomes tachyonic, which is quite typical for systems in unstable equilibrium [9,10]. This situation corresponds to an unstable electromagnetic waveguide shown schematically in Fig. 1b. We will consider the physical meaning of this result later in the Discussion section.

If we now consider a more general case of

$$\phi = Gz + Hz^2 \quad , \quad (9)$$

where G is a weak non-zero gravitational field, a similar consideration leads to a photon effective mass

$$m_{eff}c^2 = 2\hbar\left(H + \frac{3G^2}{c^2}\right)^{1/2} \quad (10)$$

The latter expression demonstrates that in the Newtonian limit a weak non-zero gravitational field leads only to a small second-order correction.

## Gravitons in the Newtonian limit

It is instructive to verify that gravitons exhibit the same behaviour as photons in the Newtonian limit of gravity. As usual, we will assume small perturbations of metric (4) of the following form:

$$g_{ik} = g_{ik}^{(0)} + \eta_{ik},\quad (11)$$

where $\eta_{ik}$ is the weak graviton field obeying the covariant wave equation [7]:

$$\eta_{ik;l}^{;l} = 0 \quad (12)$$

(where index ;l indicates the covariant derivative). For a plane gravitational wave propagating in the *x* direction the independent states of polarization are $\eta_{23}$ and $\eta_{22} = \eta_{33}$ [7]. By taking into account that based on Eq.(4)

$$\Gamma_{3k}^{k} = \frac{1}{2g}\frac{\partial g}{\partial z} \quad (13)$$

and

$$\Gamma_{33}^{3} = -\frac{2Hz}{c^2\left(1 - \frac{2Hz^2}{c^2}\right)}, \quad (14)$$

the covariant wave equation may be written as

$$\left(\Box + \frac{\partial}{\partial z}\left(\frac{g^{33}}{2g}\frac{\partial g}{\partial z}\right) + \Gamma_{33}^{3}\frac{g^{33}}{2g}\frac{\partial g}{\partial z}\right)\eta_{ik} \approx \left(\Box - \frac{4H}{c^2}\right)\eta_{ik} = 0, \quad (15)$$

where

$$\Box = \Delta - \frac{\partial^2}{c^2\partial t^2} \quad (16)$$

is the d'Alembertian. Once again, we find that for an observer in the vicinity of z=0 the *4H/c²* term plays the role of effective graviton mass squared, even though the covariant wave equation (12) is massless. Similar to photons, the effective graviton mass is positive at *H>0* , and at *H<0* the wave equation (15) becomes tachyonic.

## Effective photon mass in the Kottler–Møller spacetime

Let us now move beyond the Newtonian limit, and consider the photon wave equation in the Kottler–Møller metric

$$ds^2 = -\left(1 + \frac{za}{c^2}\right)^2 dt^2 + dx^2 + dy^2 + dz^2, \quad (17)$$

where *a* is acceleration. Based on Eq.(3), the effective dielectric permittivity and magnetic permeability in this spacetime equal

$$\varepsilon = \mu = \frac{1}{1 + \frac{za}{c^2}}, \quad (18)$$

leading to a wave equation (see Eq.(7)):

$$-\frac{\partial^2 \psi}{\partial z^2} + \frac{-\frac{\omega^2}{c^2} - \frac{a^2}{4c^4}}{\left(1 + \frac{za}{c^2}\right)^2}\psi = -k^2\psi \quad (19)$$

An accelerating observer located near the *z*=0 plain perceives photons as a tachyonic field with an effective mass

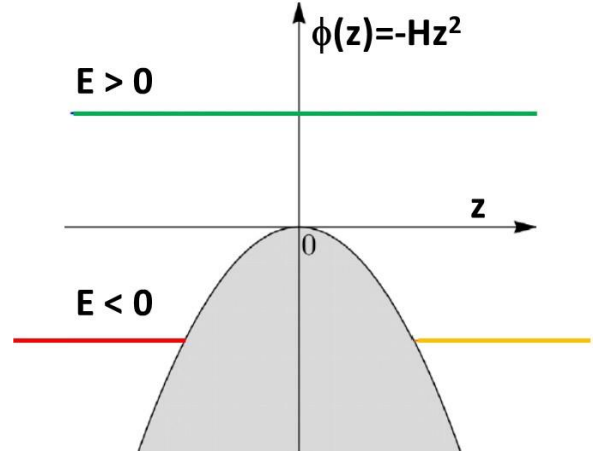

**Fig. 2.** Energy eigenstates in an inverted harmonic oscillator potential. While particles with positive energies (shown in green) may surpass the potential barrier, classical particles with negative energy (shown by orange and red lines) cannot penetrate the barrier.

$$im_{eff}c^2 = \frac{\hbar a}{2c} = \pi k_B T_U, \quad (20)$$

where $k_B$ is the Boltzmann constant. Note that the absolute magnitude of the tachyonic effective mass of photons coincides with the Unruh temperature $T_U$ (expressed in natural units).

## Discussion

A typical consequence of having the tachyonic degrees of freedom in an unstable physical system is that the system must undergo a transition to a stable equilibrium state [9,10]. In the case of Unruh effect [11], the physical vacuum transitions into a thermal state with temperature $T_U$ as perceived by an accelerating observer. The discussion above elucidates the direct connection between the Unruh effect and the appearance of tachyonic degrees of freedom in an accelerated reference frame. Note that in the equivalent electromagnetic medium situation (described by Eq.(18)) the thermal state originates from the divergencies of $\varepsilon$ and $\mu$ at the Rindler horizon located at *z=-c²/a*.

A similar situation should be expected in the case of negative gravity gradient *H<0* considered above in the Newtonian limit. The tachyonic instability of electromagnetic field near the unstable equilibrium plain should lead to the appearance of thermal radiation bath (as perceived by an observer located near *z*=0) with temperature approximately equal to the magnitude of the observed tachyonic photon mass

$$im_{eff}c^2 = k_B T_H = 2\hbar|H|^{1/2}, \quad (21)$$

where the small correction due to weak non-zero *G* is neglected. In the equivalent electromagnetic medium geometry this thermal state originates from the divergencies of $\varepsilon$ and $\mu$ at $z=\pm c/(-2H)^{1/2}$, which only happens at negative $H$ when the effective photon mass is tachyonic (see Eq.(6)). Note that the inverted harmonic oscillator potential (which is shown schematically in Fig. 1b and Fig. 2) exhibits many other similarities with the event horizon physics. It was demonstrated in [12,13] that horizons in phase space and the logarithmic phase singularity, which is a characteristic feature of Hawking radiation and Unruh effect, do appear in the inverted harmonic oscillator potential even in the classical mechanics ($c\to\infty$) limit.

Hawking radiation and Unruh effect are closely related to particle production in strong inhomogeneous gravitational fields. It is well known that gravitational tidal forces (which cannot be transformed away through adequate coordinate transformations) may be responsible for separation of the particles in a virtual pair [14]. For example, in the case of weak inhomogeneous gravitational field the local rate of massless particle production was calculated in [14] as

$$\Gamma = \frac{\alpha}{32\pi} C_{iklm} C^{iklm} \quad , \quad (22)$$

where $C_{iklm}$ is the Weyl conformal tensor, and $\alpha$ is equal to 1/60 in the case of massless scalar particles, and 1/5 in the case of photons. Near $z=0$ $C_{iklm}=R_{iklm}$ since $R_{ik}=0$. As a result, we obtain

$$\Gamma = \frac{3H^2}{10\pi c^4} \quad , \quad (23)$$

A photon pair creation must be expected when the work exercised by gravitational tidal forces over the photon wavelength $\lambda$ exceeds the energy of the photon pair, so that

$$\left(\frac{\hbar\omega}{c^2}\right) H\lambda^2 \sim \hbar\omega \quad (24)$$

This simple estimate indicates that the average energy of photons emitted by the inverted harmonic oscillator is

$$\hbar\omega = \frac{2\pi\hbar c}{\lambda} \sim k_B T_H \sim 2\pi\hbar |H|^{1/2}, \quad (25)$$

which coincides with Eq.(21) for $T_H$. Note that in the vicinity of the $z=0$ plain the photons in the pair are emitted into the $E<0$ states which move in the opposite directions and live in different half spaces. As described in detail in ref.[12], these photons are separated by a horizon in phase space. Thus, photon pair production in the vicinity of unstable equilibrium plain at $z=0$ looks quite similar to Hawking radiation near a black hole event horizon.

Admittedly, similar to Unruh effect [11], the effect described by Eqs.(21,25) is very small, unless the gravity gradients are truly enormous, and it should be very difficult to observe this effect in the experiment. For example, the observed thermal bath temperature $T_H$ would reach room temperature at $H \sim 4 \cdot 10^{26}$ s$^{-2}$. The only potential possibility to observe this effect in the experiment is to study analogue optical systems, where the effective accelerations may reach up to $10^{24}g$, as described in detail in ref. [2]. In that paper it was proposed to observe Unruh effect in a tapered optical waveguide geometry, which emulates strongly accelerated motion of photon quasi-particles. In the related work [15] it was estimated that the effective gravity gradients may reach magnitudes up to $H \sim 10^{31}$ s$^{-2}$ in the optical metamaterial waveguide geometries.

On the other hand, the prediction that an observer located in a gravity gradient $H$ near the unstable equilibrium plain at $z=0$ perceives physical vacuum as a bath of thermal radiation with temperature $k_B T_H \sim \hbar H^{1/2}$ (see Eq.(21)) constitutes yet another very interesting effect of quantum gravity, which supplements theoretical predictions of Hawking radiation and Unruh effect. Note that in the gravitational potential described by Eq. (5) $T_H$ exceeds the Unruh temperature $T_U$ at small $z$:

$$z < \frac{c}{2H^{1/2}} \quad (26)$$

Moreover, unlike Unruh temperature, the magnitude of $T_H$ is unaffected in the non-relativistic $c\to\infty$ limit, which makes this effect considerably different from Hawking and Unruh temperature, which disappear in the $c\to\infty$ limit. The latter feature arises from the fact that an alternative way to derive the existence of thermal background at the unstable equilibrium $z=0$ position is to consider surface gravity at the two distant horizons located at $z=\pm c/(-2H)^{1/2}$.
The surface gravity is

$$a = c(2H)^{1/2} \quad , \quad (27)$$

and the Hawking temperature originating from these horizons coincides with Eqs. (21,25).

We should also note that the existence of gravity-induced effective photon mass (even extremely small) is important, since it considerably narrows down the number of reference frames which may be counted as truly inertial. Simple elimination of the gravitational field (but not its gradients) is clearly insufficient, since observers located near $z=0$ and experiencing no gravitational field may still experience "immersion" into a thermal radiation bath having different temperatures, and these observers may also measure different effective photon masses near the $z=0$ plain. Note however that the described emergence of the effective photon mass does not lead to a violation of gauge symmetry and charge conservation, since at the fundamental level the electromagnetic field equations and the wave equation for gravitational waves (Eq.(12)) remain massless.